\documentclass[leqno,a4paper,11pt]{article}

\usepackage[colorlinks,
linkcolor=blue,
urlcolor=blue,
citecolor=red]{hyperref}
\usepackage{amsthm}
\usepackage{amsmath}
\usepackage{amssymb}
\usepackage[normalem]{ulem}

\usepackage{mathtools}
\mathtoolsset{showonlyrefs}

\newcommand{\Con}{\ensuremath{\mathcal{C}}}

\newcommand{\mb}[1]{\ensuremath{\mathbb{#1}}}

\newcommand{\R}{\mb{R}}

\newfont{\bl}{msbm10 scaled \magstep2}

\newcommand{\beq}{\begin{equation}}
\newcommand{\eeq}{\end{equation}}

\newcommand{\notmid}{\mid\kern-0.5em\not\kern0.5em}

\newenvironment{pr}{\begin{proof}[\textbf{Proof:}] \ }{\end{proof}}
\newtheorem{thm}{Theorem}[section]
\newtheorem{lem}[thm]{Lemma}

\newtheorem{cor}[thm]{Corollary}
\theoremstyle{definition}
\newtheorem{defi}[thm]{Definition}
\theoremstyle{remark}
\newtheorem{ex}[thm]{Example}

\newcommand{\LLS}{Lorentzian length space }
\newcommand{\LLSn}{Lorentzian length space}
\newcommand{\LpLS}{Lorentzian pre-length space }
\newcommand{\LpLSn}{Lorentzian pre-length space}
\newcommand{\Xll}{\ensuremath{(X,d,\ll,\leq,\tau)} }

\newcommand{\LL}{\mb{L}}

\usepackage{enumitem}

\usepackage{framed}

\usepackage[most]{tcolorbox}

\renewcommand{\Xll}{$(X,\ell)$ }

\renewcommand{\labelenumi}{(\roman{enumi})}
\renewcommand\theenumi\labelenumi

\newtcolorbox{attentionbox}[1][]{%
  enhanced jigsaw,
  sharp corners,
  colframe={red},
  underlay={%
    \path[draw=none] (interior.south west) rectangle node[white]{{\color{red} \Huge !}} ([xshift=-10mm]interior.north west);
    },
  #1
}

\title{Curvature bounds, regularity and inextendibility of spacetimes}
\author{Tobias Beran\thanks{{\href{mailto:BeranT@cardiff.ac.uk}{\normalfont\ttfamily BeranT@cardiff.ac.uk}} School of Mathematics, Cardiff University}, John Harvey\thanks{{\href{mailto:HarveyJ13@cardiff.ac.uk}{\normalfont\ttfamily HarveyJ13@cardiff.ac.uk}}, School of Mathematics, Cardiff University}, Clemens S\"amann\thanks{{\href{mailto:clemens.saemann@univie.ac.at}{\normalfont\ttfamily clemens.saemann@univie.ac.at}}, Faculty of Mathematics, University of Vienna.}\ \thanks{Corresponding author}}

\usepackage{makeidx}
\makeindex

\begin{document}

\maketitle

{\abstract{We provide a completely new relation between curvature bounds and definiteness of the causal character of maximizers by exploiting the robust notion of synthetic curvature. This enables us to relate low-regularity inextendibility of spacetimes to unboundedness of curvature --- which is at present unattainable using classical methods --- thereby strengthening and complementing the results of \cite{GKS:19} significantly.
\vskip 1em

\medskip

}}


\section{Introduction}
A long-standing problem in general relativity is the development of a precise definition of a singularity. From a mathematical point of view Penrose suggested that a singularity is an incomplete causal geodesic, i.e., the world line of an observer that suddenly stops existing (or suddenly comes into existence). This definition has the advantage of being rigorously defined and Penrose, and then Hawking, used this for their famous \emph{singularity theorems} \cite{Pen:65, Haw:67}. The disadvantage is that there is no clear implication to physical singularities, i.e., a blow-up of curvature or the existence of a horizon. To partly remedy this one should only consider \emph{inextendible} spacetimes, i.e., spacetimes that cannot be viewed as proper subsets of another spacetime of the same dimension (which is called \emph{the extension}). 

However, should one consider smooth extensions only or should one allow extensions of low regularity? Both from a physical perspective and from the mathematical one, the latter is appropriate as low regularity extensions can still be non-singular. Sbierski recently invigorated this line of research by establishing the $\Con^0$-inextendibility of Schwarzschild spacetime \cite{Sbi:18, Sbi:18b} and e.g.\ \cite{GLS:18, Sbi:22, Sbi:25}. The $\Con^0$-inextendibility of FLRW-spacetimes has also been studied \cite{GL:17}, as have been boundary extensions \cite{Sbi:24, GvBS:24}.

Recently there has been substantial progress in a \emph{synthetic approach} to Lorentzian geometry and general relativity. In particular, it became possible to use curvature (bounds) in highly non-smooth situations (e.g.\ $\Con^0$-spacetimes) or spaces that are not manifolds (discrete spaces, causal sets \cite{Sur:25, Bra:26} etc.). See \cite{CM:22, Sae:24, McC:25, Bra:25} for an overview of this line of research. The main idea is to describe curvature in a robust way by using just (Lorentzian) distances (i.e., time separation) and volumes. Such a perspective proved to be remarkably fruitful in the Riemannian case and led to the development of the fields of \emph{metric geometry} (e.g.\ Alexandrov and $\mathsf{CAT}(k)$-spaces \cite{BH:99, BBI:01, AKP:24}) and \emph{metric measure geometry} (e.g.\ $\mathsf{(R)CD}$-spaces \cite{LV:09, Stu:06c, Vil:09}). The former generalizes sectional curvature bounds to metric spaces; the latter generalizes Ricci curvature bounds to metric spaces with a reference measure (i.e., the volume). In the Lorentzian case \emph{timelike sectional curvature bounds} have been introduced in \cite{KS:18} and \emph{timelike Ricci curvature bounds} in \cite{CM:24a} building on \cite{McC:20, MS:23}. In particular, for smooth ($\Con^2$ is enough), strongly causal spacetimes bounds on the timelike sectional curvature are equivalent to the synthetic curvature bounds via Lorentzian distance (time separation) comparison \cite{BKOR:25}, while for timelike Ricci curvature the equivalence for smooth, globally hyperbolic spacetimes has been established in \cite{McC:20,MS:23}.

Now, the significant advantage of having a synthetic notion of curvature is that one has a notion of curvature at one's disposal even for non-smooth extensions of classical smooth spacetimes, e.g.\ $\Con^0$-extensions. Restricting to classical techniques does not at present allow us to investigate whether low regularity extensions have bounded curvature. A proof-of-concept of this was established in \cite{GKS:19} but there has not been any follow-up since then. The goal of this article is to revitalize this line of research. 

We give a completely novel result showing that \emph{regularity}, i.e., that maximizers are either timelike or null, follows from timelike or causal curvature bounded below under mild causality assumptions. 
From this we prove, among other results, that a smooth and strongly causal spacetime which is timelike geodesically complete cannot be extended as a weakly normal \LpLS having causal or timelike curvature bounded below.

Note that these results contrast to \cite{GKS:19} where \emph{upper} curvature bounds and the full strength of \emph{\LLSn s} were used. Here we only work with \emph{\LpLSn s} (see Section \ref{sec-lpls-cbs} for the precise definitions). Furthermore, a correction to \cite[Prop.\ 4.15]{KS:18}\footnote{See the updated arXiv-version at \url{doi.org/10.48550/arXiv.1711.08990}.}, made after the publication of \cite{GKS:19}, implies that the main result of \cite{GKS:19} depends on the rather unnatural assumption of local timelike geodesic connectedness. Here, we require only mild and natural causality assumptions.

\section{\LpLSn s and curvature bounds}\label{sec-lpls-cbs}
We briefly recall the basics of \LpLSn s \cite{KS:18} and take new developments into account. In particular, there is no need to fix a particular background metric. It is enough to demand that the topology is metrizable and finer than the chronological topology, i.e., the one generated by the chronological futures and pasts (see below). This point of view has recently been taken in \cite{MS:25}. Moreover, it is enough to consider an extended time separation function and derive the timelike and causal relations from it. This has been advocated by McCann \cite{McC:24} and used several times since then, see e.g.\ \cite{BMcC:23, BBCGMORS:24}. Moreover there are other approaches e.g.\ the bounded Lorentzian metric spaces \cite{MS:24, BMS:25} of Minguzzi--Suhr(\nobreakdash--Bykov) or the null distance of Sormani--Vega \cite{SV:16}.

Let $X$ be a set and let $\ell\colon X\times X\rightarrow \{-\infty\}\cup[0,\infty]$ satisfy the \emph{reverse triangle inequality}, i.e.,
 \begin{equation}\label{eq-rev-tri-ine}
  \ell(x,y) + \ell(y,z) \leq \ell(x,z)\qquad \forall x,y,z\in X\,,
 \end{equation}
employing the convention that $-\infty + \infty = \infty + (-\infty) = -\infty$. Then, the \emph{timelike} $\ll$ and \emph{causal $\leq$ relations} are $\ll:=\ell^{-1}((0,\infty])$ and $\leq:=\ell^{-1}([0,\infty])$. We write $x<y$ for $x\leq y$ and $x\neq y$. Analogously to the classical spacetime case, the \emph{chronological} and \emph{causal future (past)} of $x\in X$ are
 \begin{align*}
  I^+(x)&:=\{y\in X: x\ll y\}\,, \qquad I^-(x):=\{y\in X: y\ll x\}\,,\\
  J^+(x)&:=\{y\in X: x\leq y\}\,, \qquad J^-(x):=\{y\in X: y\leq x\}\,,
 \end{align*}
and the \emph{chronological} and \emph{causal diamonds} are $I(x,y):= I^+(x)\cap I^-(y)$, $J(x,y):=J^+(x)\cap J^-(y)$. Having the chronological relation at hand we define the \emph{chronological topology} as the topology on $X$ generated by the sub-base of chronological futures and pasts $I^\pm(x)$ ($x\in X$). Then a \LpLS is defined as follows.

\begin{defi}[\LpLSn]\label{defi-lpls}
 Let $X$ be a set, $\ell\colon X\times~X\rightarrow \{-\infty\}\cup[0,\infty]$ as above and $\ell(x,x)\geq 0$ for all $x\in X$. If $X$ is given a metrizable topology that is finer than the chronological one, then \Xll is a \emph{\LpLSn}. Also, $\tau:=\max(0,\ell)$ is the \emph{time separation function} and $\ell$ is called the \emph{extended time separation function}. 
\end{defi}

Having now the causal relations at hand, we can define a \emph{timelike} or \emph{causal} curve as a continuous curve that is totally ordered by $\ll$ and $\leq$, respectively. Then the length of a causal curve $\gamma\colon[a,b]\rightarrow X$ is
\begin{align}
 L_\tau(\gamma):=\inf \Bigl\{ \sum_{i=1}^N \tau(\gamma(t_{i-1}),\gamma(t_i)): a=t_0< t_1<\ldots t_N=b\Bigr\}\,.
\end{align}
A \emph{(causal) maximizer} is a causal curve whose length saturates the time separation between its endpoints, i.e., $\tau(\gamma(a),\gamma(b)) = L_\tau(\gamma)$. We also write $[x,y]$ for a choice of maximizer from $x$ to $y$. Moreover, $\tau$ (and hence $(X,\ell)$) is \emph{intrinsic} if
\begin{align}
 \ell(x,y) = \sup \bigl\{L_\tau(\gamma): \gamma \text{ causal from } x \text { to } y\bigr\}\,.
\end{align}

A \LpLS is \emph{future distinguishing} if $I^+(x)=I^+(y)$ implies $x=y$ for all $x,y\in X$. Analogously, it is called \emph{past distinguishing} if $I^-(x)=I^-(y)$ implies $x=y$. It is called \emph{distinguishing} if it is both future and past distinguishing, cf.\ \cite[Def.\ 4.59]{Min:19b}. It is \emph{locally (past/future) distinguishing} if for each neighbourhood $U$ of $x$ and $y$, the same holds when intersecting with $U$. We call $X$ \emph{not locally timelike isolating} if for all neighbourhoods of all $x\in X$, both $I^+(x)\cap U$ and $I^-(x)\cap U$ are non-empty \cite[Def.\ 2.7]{Rot:23}. These causality assumptions will be used in Section \ref{sec-reg} to obtain regularity from lower curvature bounds.

While locally distinguishing obviously implies distinguishing, the converse is also true for \emph{timelike path-connected} spaces, i.e., spaces where all $x\ll y$ can be connected by a timelike curve. For an example of a space that fails to be distinguishing but is not locally timelike isolating consider the Lorentz cylinder, i.e., $\mathbb R\times[0,1]$ mod $(t,0)\sim(t+1,1)$. For an example that is locally distinguishing but locally timelike isolating it suffices to consider a closed slim diamond in Minkowski spacetime, e.g.\ $\{(x,t) : t\geq0, |x| \leq t/2\} \cap \{(x,t) : t\leq 1, |x| \leq (1-t)/2\}$.

A \LpLS is \emph{strongly causal} if its (metrizable) topology is generated by the sub-base of chronological diamonds $I(x,y)$ for $x,y\in X$. A strongly causal \LLS is distinguishing by \cite[Prop.\ 3.17]{ACS:20}. However, for \LpLSn s the situation is a bit more subtle. 
In any case, in this paper the results establishing regularity rely directly on the distinguishing condition rather than on strong causality.

Finally, a \LpLS is called \emph{regular} if every maximizer is either timelike or null.


\medskip

Moreover, we introduce here a new notion, that of \emph{weakly normal neighbourhoods}, generalizing normal (or convex) neighbourhoods for smooth spacetimes and localizing neighbourhoods of \cite{KS:18}. We first recall the definition of a localizing neighbourhood. As we will see, we do not need the full power of localizability to prove Lemma \ref{lem:localizingImpliesWeaklyNormal}.

\begin{defi}[Localizable spaces]
Let \Xll be a Lorentz\-ian pre-length space. A neighbourhood $U$ of $p$ is a \emph{localizing neighbourhood} if there is a metric $d$ compatible with the topology such that the following holds:
\begin{itemize}
\item $U$ is a $d$-compatible neighbourhood, i.e., there is a uniform upper bound on the $d$-length of causal curves in $U$,
\item $U$ is not timelike isolating, i.e., $I^\pm(p)\cap U \neq \emptyset$,
\item $U$ as a space is strongly causal,
\item local maximizers in $U$ exist, i.e., for all $x,y\in U$ with $x<y$ there is a causal curve from $x$ to $y$ within $U$ which is not shorter than any other causal curve from $x$ to $y$ within $U$,
\item the local time separation function $\omega_U:U\times U\to [0,\infty)$, defined by $\omega_U(p,q)=L_\tau(\gamma)$ for $p\leq q$ and $\gamma$ the local maximizer in $U$, is finite and continuous.
\end{itemize}
Furthermore, we call \Xll \emph{localizable} if every point has a localizing neighbourhood.
\end{defi}

By comparison, weak normality makes fewer assumptions, though these assumptions are stricter or not directly comparable to those of localizability.

\begin{defi}[Weakly normal neighbourhood]
 Let \Xll be a Lorentz\-ian pre-length space and $p\in X$. A neighbourhood $U$ of $p$ is \emph{weakly normal}~if
 \begin{enumerate}
  \item the time separation function $\tau$ is finite on $U$, i.e., $\tau\rvert_{U\times U}<\infty$,
  \item maximizers exist, i.e., for all $x,y\in U$ with $x<y$ there is a causal maximizer from $x$ to $y$,
  \item and it is not timelike isolating.
 \end{enumerate}
 Furthermore, we call \Xll \emph{weakly normal} if every point has a weakly normal neighbourhood.
\end{defi}

Note that the maximizer required to exist by point (ii) need not be contained in $U$. One could also demand this or demand additionally that it is only maximizing among all curves in $U$ (similar to localizing neighbourhoods, cf. \cite[Def.\ 3.16]{KS:18}) but this will not be needed for our purposes. Also note all previous notions of curvature comparison neighbourhoods demand the existence of maximizers and $\tau$ finite-valued there, cf.\ Definition \ref{defi-cbb} and \cite{KS:18, BS:23, BKR:24}. Thus all spaces with curvature bounds (except the one for discrete spaces where existence of maximizers is not required) are weakly normal. Moreover, we have the following compatibility result.

\begin{lem}[Intrinsic, strongly causal and localizing implies weakly normal]\label{lem:localizingImpliesWeaklyNormal}
 Let \Xll be a strongly causal and localizable \LpLS that is intrinsic, then it is weakly normal.
\end{lem}
\begin{pr}
 We can follow the argument of \cite[Lem.\ 4.3]{GKS:19}, which we provide here for convenience: Let $p\in X$ and $\Omega$ a localizing neighbourhood of $p$ with local time separation function $\omega$, which is finite-valued. Strong causality implies that there is a neighbourhood $U$ of $p$ with $U\subseteq \Omega$ such that all causal curves with endpoints in $U$ are contained in $\Omega$ (via \cite[Lem.\ 2.38,(iii)]{KS:18}). Let $x,y\in U$ with $x <y$, as $\Omega$ is localizing there is a causal curve $\sigma$ from $x$ to $y$ that is maximal in $\Omega$ with $L_\tau(\sigma)=\omega(x,y)$. By the properties of $U$ every causal curve connecting $x$ and $y$ is contained in $\Omega$. Thus, $\sigma$ is maximal in $X$ and so, as $\tau$ is intrinsic, we have $\tau(x, y) = L_\tau(\sigma) =\linebreak \omega(x, y)<\infty$.
\end{pr}
In fact, the lemma implies that every strongly causal \LLSn, and hence every strongly causal spacetime is \mbox{weakly normal}.
\smallskip

At this point we recall curvature bounds for \LpLSn s, see \cite{KS:18, BKR:24} for more details. We will focus on \emph{triangle comparison} and the so-called \emph{four-point condition}. The two-dimensional Lorentzian model spaces of constant curvature $k\in\R$ (see e.g.\ \cite{ONe:83}) are $\LL^2(k):= \tilde S^2_1(\frac{1}{\sqrt{k}})$ for $k>0$, $\LL^2(0):=\R^2_1$, and $\LL^2(k):= \tilde H^2_1(\frac{1}{\sqrt{-k}})$ for $k<0$, which have diameter $D_k:=\frac{\pi}{\sqrt{-k}}$ if $k<0$ and $D_k:=\infty$ otherwise. Here $\tilde S^2_1(r)$ is the simply connected covering manifold of the two-dimensional Lorentzian pseudosphere of radius $r>0$ ($r=1$ is de Sitter space), $\R^2_1$ is two-dimensional Minkowski spacetime and $\tilde H^2_1(r)$ is the simply connected covering manifold of two-dimensional Lorentzian pseudo-hyperbolic space ($r=1$ is anti-de Sitter space). 

A \emph{timelike triangle} is a triple $\Delta(x,y,z)$ of timelike related points $x\ll y \ll z$ 
with a choice of maximizing sides. A \emph{comparison triangle} of $\Delta(x,y,z)$ is a timelike triangle $\Delta(\bar x,\bar y,\bar z)$ in $\LL^2(k)$ which has the same side lengths as $\Delta(x,y,z)$. For every timelike triangle $\Delta(x,y,z)$ with $\tau(x,z)<D_k$, there is a comparison triangle in $\LL^2(k)$ and this is unique up to isometry of $\LL^2(k$), cf.\ \cite[Lem.\ 2.1]{AB:08}.

\begin{defi}[Triangle comparison]\label{defi-cbb}
 Let \Xll be a \LpLS and $k\in\R$. A \emph{$\geq k$-comparison neighbourhood} (in the sense of triangle comparison) is an open set $U\subseteq X$ such that
 \begin{enumerate}
  \item $\tau$ is continuous on $(U\times U)\cap\tau^{-1}([0,D_k))$ and this set is open,
  \item $U$ is $D_k$-geodesic, i.e., for all $x,y\in U$ with $0<\tau(x,y)<D_k$ there is a maximizer from $x$ to $y$ in $U$.
  \item Let $\Delta(x,y,z)$ be a timelike triangle in $U$ with $\tau(x,z)<D_k$ and let $p,q$ be two points on its sides. Let $\Delta(\bar x,\bar y,\bar z)$ be a comparison triangle in $\LL^2(k)$ of $\Delta(x,y,z)$ and $\bar p, \bar q$ on the corresponding sides with equal time separations from the vertices as $p,q$. Then we have $\tau(p,q)\leq \bar\tau(\bar p,\bar q)$.
 \end{enumerate}
Then \Xll has \emph{timelike curvature bounded below (TLCBB) by $k$ (in the triangle comparison sense} if every point has a $\geq k$-comparison neighbourhood (in the triangle comparison sense).
\end{defi}

Instead of comparing timelike (or causal) triangles to comparison triangles in $\LL^2(k)$ we can also use \emph{four-point configurations}, which have the advantage that we do not need the existence of maximizers.

 \begin{itemize}
  \item A quadruple $(y,x,z_1,z_2)\in X^4$ is a \emph{causal future four-point configuration} if $y\ll x\leq z_i$ for $i=1,2$. Similarly, a \emph{causal past four-point configuration} is a quadruple $(z_2,z_1,x,y)\in X^4$ such that $z_i\leq x\ll y$ for $i=1,2$.
  \item For a causal future four-point configuration $(y,x,z_1,z_2)$ and $k\in~\R$, a \emph{four-point comparison configuration} in $\LL^2(k)$ is a quadruple $(\bar y,\bar x,\bar z_1,\bar z_2)\in \LL^2(k)^4$ such that
  \begin{enumerate}
   \item $\tau(y,x)=\bar\tau(\bar y,\bar x)$\,,
   \item $\tau(y,z_i) = \bar\tau(\bar y,\bar z_i)$ ($i=1,2$)\,,
   \item $\tau(x,z_i) = \bar\tau(\bar x,\bar z_i)$ ($i=1,2$)\,, and
   \item $\bar z_1$, $\bar z_2$ lie on opposite sides of the line through $\bar y$, $\bar x$.
  \end{enumerate}
  \item Similarly, one defines a four-point comparison configuration for a causal past four-point configuration.
 \end{itemize}

Using the four-point configurations one can define synthetic causal sectional lower curvature bounds in the form of the \emph{four-point condition} as introduced by Beran--Kunzinger--Rott \cite[Def.\ 4.6]{BKR:24} following an analogous construction in the positive signature case, cf.\ e.g.\ \cite{AKP:24}. Here for brevity we only recall one condition in full.
\begin{defi}[Four-point condition]
 Let \Xll be a \LpLS and $k\in\R$. A \emph{$\geq k$-comparison neighbourhood} (in the sense of the strict causal four-point condition) is an open set $U\subseteq X$ such that for every causal four-point configuration $(y,x,z_1,z_2)$ in $U$ with $\tau(y,z_i)<D_k$ ($i=1,2$) and its comparison configuration $(\bar y,\bar x,\bar z_1, \bar z_2)$ in $\LL^2(k)$ one has 
  \begin{equation}\label{eq-fu-fou-pt}
   \tau(z_1,z_2)\geq \bar\tau(\bar z_1,\bar z_2)\quad \text{ and } \quad \bar z_1 \leq \bar z_2 \Rightarrow z_1 \leq z_2\,.
  \end{equation}
  Moreover, for every causal past four-point configuration $(z_2,z_1,x,y)$ in $U$ with $\tau(z_i,y)<D_k$ ($i=1,2$) and its comparison configuration $(\bar z_2, \bar z_1,\bar x,\bar y)$ in $\LL^2(k)$ one has 
  \begin{equation}
   \tau(z_2,z_1)\geq \bar\tau(\bar z_2,\bar z_1)\quad \text{ and } \quad \bar z_2 \leq \bar z_1 \Rightarrow z_2 \leq z_1\,.
  \end{equation}
We say that \Xll has \emph{causal curvature bounded below by $k$} (CCBB) if $X$ can be covered by $\geq k$-comparison neighbourhoods (in the strict causal four-point sense).
\end{defi}

The \emph{timelike} (TLCBB) and the \emph{(non-strict) causal four-point condition} are given by, for future configurations, dispensing with the implication $\bar z_1 \leq \bar z_2 \Rightarrow z_1 \leq z_2$ and assuming additionally that $z_1 \leq z_2$, and in the timelike case that $y \ll x \ll z_1, z_2$. Similar modifications are needed for past configurations. Also note that for all these four-point conditions we drop the assumption that $\tau$ is continuous (and finite) in the comparison neighbourhood as compared to \cite{BKR:24}, though we will need it for Theorem \ref{thm-tlfpcbb-reg}.

The four-point conditions are equivalent to the other synthetic causal and timelike sectional curvature bounds under very general assumptions, and hence to smooth timelike sectional curvature bounds, see \cite[Thm.\ 5.1]{BKR:24} and \cite{BKOR:25} --- in particular the graphic visualization on p.\ 38 (Figure 4) of \cite{BKR:24} --- but note that regularity is one of these assumptions.

\section{Curvature and Regularity}\label{sec-reg}
We now establish that a lower curvature bound implies regularity. This complements \cite[Prop.\ 4.15]{KS:18} for curvature bounded above but dispenses with the unnatural assumption of local timelike geodesic connectedness (see the corrected arXiv-version {\footnotesize \url{doi.org/10.48550/arXiv.1711.08990}}).

We show this in three different settings. First we use causal curvature bounded below, in the strict causal 4-point sense. Then we use timelike curvature bounded below, both in the four-point sense and in the triangle comparison sense. 

Curvature comparison conditions related to angles and hinges require regularity as part of the definition, as does the $\tau$-convexity definition. The results here show that conditions related to triangle comparison and four-point conditions force regularity under the mild causality assumptions of not locally timelike isolating and locally distinguishing. It now follows that these assumptions can replace regularity in the equivalence proofs of the different characterisations of lower curvature bounds from \cite{BKR:24}. 
These local causality conditions can be replaced with weaker global conditions if the spaces are assumed to have global curvature bounds.

\begin{thm}[CCBB implies regularity]\label{thm-ccbb-reg}
A locally distinguishing \LpLS \Xll that has causal curvature bounded below (CCBB) (in the strict causal 4-point sense) is regular.
\end{thm}
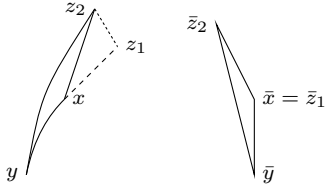
\begin{figure}
\centering\scriptsize
\begin{tikzpicture}
    \draw (0,0)  .. controls (0.1,0.5) and (0.3,0.8) ..  (0.5,1);
    \draw (0,0)  .. controls (0.2,1.1) ..  (0.9,2.2);
    \draw[dash pattern=on 2pt off 2pt] (0.5,1) -- (1.2,1.7);
    \draw (0.5,1) -- (0.9,2.2);
    \draw[dash pattern=on 1pt off 1pt] (0.9,2.2) -- (1.2,1.7);
    \draw (0,0) node[anchor=east]{$y$} (0.5,1) node[anchor=west]{$x$} (1.2,1.7) node[anchor=west]{$z_1$} (0.9,2.2) node[anchor=east]{$z_2$};
    \begin{scope}[xshift=3cm]
        \draw (0,0)--(0,1) -- (-0.5,2) -- cycle;
        \draw (0,0) node[anchor=west]{$\bar y$} (0,1) node[anchor=west]{$\bar x=\bar z_1$} (-0.5,2) node[anchor=east]{$\bar z_2$};
    \end{scope}
\end{tikzpicture}
\caption{Proof of Theorem \ref{thm-ccbb-reg}: The dashed line from $x$ to $z_1$ represents a null portion of the maximizer $[y,z_1]$. The null edge $[x,z_1]$ of the degenerate triangle with vertices $y$, $x$, $z_1$ collapses in the comparison configuration due to the regularity of the model space. It follows that $x$ and $z_1$ cannot be locally distinguished by their futures.}
\label{fig-ccbb-reg}
\end{figure}
\begin{pr}
Suppose towards a contradiction that $\gamma\colon[0,1]\rightarrow X$ is a maximizer that is not null but contains a null relation $\gamma(t_0)\not\ll\gamma(t_1)$, and without loss of generality assume $t_0>0$ and $t_0$ is minimal with this property. By the push-up property we have that the segment $\gamma|_{[t_0,t_1]}$ is null, and as $\gamma$ is a maximizer, $0<\tau(\gamma(t_0-\varepsilon),\gamma(t_1))=\tau(\gamma(t_0-\varepsilon),\gamma(t_0))+\tau(\gamma(t_0),\gamma(t_1))=\tau(\gamma(t_0-\varepsilon),\gamma(t_0))$ for all small $\varepsilon>0$. Thus, in a small enough $\geq k$-comparison neighbourhood $U$ of $x:=\gamma(t_0)$, we find the points $y:=\gamma(t_0-\varepsilon)\ll x\not\ll z_1:=\gamma(t_0+\varepsilon)$.

We now show that $I^+(x)\cap U=I^+(z_1)\cap U$, contradicting the assumption that the space is locally distinguishing. The inclusion $I^+(z_1)\subseteq I^+(x)$ follows by $x\leq z_1$ and the reverse triangle inequality. For the other inclusion, let $z_2\in I^+(x) \cap U$. 
Let $(\bar y, \bar x, \bar z_1,\bar z_2)$ be a four-point comparison
configuration in $\LL^2(k)$, where $k\in\R$ is the causal curvature bound. (See Figure \ref{fig-ccbb-reg}.)

As $\bar\tau(\bar x,\bar z_1)=0$, we know that either $\bar x$ is null related to $\bar z_1$ or $\bar x=\bar z_1$. In the former case we have the contradiction
\begin{align}
 \tau(y,x) =\bar\tau(\bar y,\bar x) + \bar\tau(\bar x,\bar z_1) < \bar\tau(\bar y,\bar z_1) = \tau(y,z_1) = \tau(y,x)\,,
\end{align}
as the broken timelike geodesic $[\bar y,\bar x]\cup [\bar x,\bar z_1]$ has to have strictly smaller length than the timelike geodesic $[\bar y,\bar z_1]$ and $\gamma$ is maximizing from $y$ to $z_1$.
Note that as $y,x,z_1$ lie on a maximizer, so do $\bar y, \bar x, \bar z_1$. As $\tau(x,z_1)=0<\tau(y,x)$ and $\LL^2(k)$ is regular, we have to have $\bar x=\bar z_1$.

The strict causal 4-point comparison (CCBB) now gives
\begin{align}\label{eq-tlcbb}
  \tau(z_1,z_2) \geq \bar\tau(\bar z_1,\bar z_2)=\bar\tau(\bar x,\bar z_2)=\tau(x,z_2)>0\,,
\end{align}
making $z_2\in I^+(z_1)$. This holds for all $z_2\in I^+(x)\cap U$, thus $I^+(x)\cap U=I^+(z_1)\cap U$, contradicting local distinguishing. 
Moreover, note that we have $\tau(z_1,z_2)\leq\tau(x,z_2)$ by the reverse triangle inequality, proving that $\tau(x,.)=\tau(z_1,.)$ on $U$ (which will be used in later arguments).
\end{pr}

Note that in the above proof we do not make use of the fact that  $\bar z_1 \leq \bar z_2 \Rightarrow z_1 \leq z_2$. However, we do use the fact that the strict causal sense does not require the causal relationship $z_1 \leq z_2$. It is this additional strength of the strict condition that means we do not require the assumption that $X$ is not locally timelike isolating, which is used in the rest of this section.

Two weaker versions of the four-point curvature condition are curvature bounded below in the timelike four-point sense and in the (non-strict) causal four-point sense. Here we show that for these conditions, regularity is implied under the additional assumption that the space is not locally timelike isolating. 

\begin{thm}[4-point TLCBB implies regularity]\label{thm-tlfpcbb-reg}
Let \Xll be a \LpLS that is locally distinguishing and not locally timelike isolating.
If $X$ has timelike curvature bounded below (TLCBB) (in the timelike 4-point sense) and has $\tau$ continuous within small enough comparison neighbourhoods, then $X$ is regular.
\end{thm}
\begin{pr}
As in the previous proof, we can suppose towards a contradiction that there is a maximizer $\gamma\colon[0,1]\rightarrow X$ with $\gamma(0)=:y\ll \gamma(t_0) =: x \not\ll\gamma(1)$ contained in a small enough $\geq k$-comparison neighbourhood $U$. We want to show that $I^+(x)\cap U=I^+(\gamma(1))\cap U$. As before, $I^+(\gamma(1))\subseteq I^+(x)$ is immediate.

Let $V_n \subset U$ be a nested sequence of neighbourhoods of $\gamma(1)$ such that $\bigcap_n V_n = \{\gamma(1)\}$.
Let $z_1^{(n)} \in I^+(\gamma(1)) \cap V_n$. Such a point exists since $X$ is not locally timelike isolating.
Let $z_2\in I^+(\gamma(1))\cap~U$. Then for $n$ large enough we have $V_n\subseteq I^-(z_2)$ and thus $z_2$ will be in $I^+(z_1^{(n)})$. 

Let $(\bar y,\bar x,\bar z_1^{(n)},\bar z_2)$ be a four-point comparison configuration in $\LL^2(k)$, where $k\in\R$ is the timelike curvature bound.
Since comparison configurations vary continuously with the side lengths and as $\tau$ is continuous in $U$, we have that as $n \to \infty$ the comparison configurations converge. As in the proof of Theorem \ref{thm-ccbb-reg}, in the limit we have that $\bar z_1^{(n)} \to \bar x$.

Now by the reverse triangle inequality and curvature comparison, we have 
\begin{align}
 \tau(x,z_2) &\geq \tau(x,z_1^{(n)}) + \tau(z_1^{(n)},z_2) \geq \tau(z_1^{(n)},z_2) \\&\geq \bar\tau(\bar z_1^{(n)},\bar z_2) = \bar\tau(\bar x,\bar z_2) - \varepsilon(V_n) = \tau(x,z_2) - \varepsilon(V_n)\,,
\end{align}
where as $n \to \infty$ and $\bar z_1^{(n)} \to \bar x$, we have $\varepsilon(V_n) \to 0$.
Continuity of $\tau$ yields $\tau(z_1^{(n)},z_2)\to\tau(\gamma(1),z_2)$, hence in the limit we obtain
\begin{align}
 \tau(x,z_2) \geq \tau(\gamma(1),z_2) \geq \tau(x,z_2)\,.
\end{align}

Thus we have shown that $\tau(x,z) = \tau(\gamma(1),z)$ for all $z\in I^+(\gamma(1))\cap~U$. 

More generally, let $z \in I^+(x) \cap U$. Set $t_1 = \sup\{t : \tau(\gamma(t),z) > 0\}$. In particular, $t_1 > t_0$ and for all $t \in [t_0,t_1)$, $z \in I^+(\gamma(t))$. By the above $\tau(x,z) = \tau(\gamma(t),z)$ and hence, as a function of $t$, $\tau(\gamma(t),z)$ is constant on $(t_0,t_1)$. Since $\tau(\gamma(t_0),z)>0$ and $\tau$ is continuous in $U$ we deduce that $t_1 = 1$, $\tau(\gamma(t),z)>0$ for all $t \in [t_0, 1]$, and hence $z \in I^+(\gamma(1))$. Therefore $x$ and $\gamma(1)$ can not be distinguished from the future within $U$, violating the assumption that \Xll is locally distinguishing. 
\end{pr}

\begin{cor}[Non-strict 4-point CCBB implies regularity]\label{cor-tlfpcbb-reg}
Let \Xll be a \LpLS that is locally distinguishing and not locally timelike isolating.
If $X$ has causal curvature bounded below (in the non-strict causal 4-point sense) and has $\tau$ continuous within small enough comparison neighbourhoods, then $X$ is regular.
\end{cor}

Looking at \cite[Thm.~5.1]{BKR:24}, we see that the different versions of curvature comparison are in general only equivalent under the assumption of regularity. 
The six different versions of triangle comparison are equivalent without any additional assumptions, and hence the following result applies to all forms of triangle comparison lower curvature bounds.

\begin{thm}\label{tlcbb-reg}
A locally distinguishing \LpLS \Xll with TLCBB in the triangle comparison sense is regular.
\end{thm}
\begin{pr}
Let $\gamma$, $y = \gamma(0)$, $x = \gamma(t_0)$ and $z_1 = \gamma(1)$ be as in Theorem \ref{thm-ccbb-reg}.
Let $U$ be a $\geq k$-comparison neighbourhood for $x$ and, as in the proof of Theorem \ref{thm-ccbb-reg}, we can assume that $\gamma$ lies entirely in $U$. We claim that $I^+(x) \cap U=I^+(z_1) \cap U$, contradicting the assumption that the space is locally distinguishing: $I^+(z_1)\subseteq I^+(x)$ is immediate. For now let $z_2\in I^+(z_1) \cap U$.
Consider the timelike triangle $\Delta(y, z_1, z_2)$ with comparison point $x\in[y, z_1]$. 
As before, we have in the comparison configuration that $\bar x = \bar z_1$.
Curvature comparison implies $\tau(x,z_2)\leq\bar\tau(\bar x,\bar z_2)=\bar\tau(\bar z_1,\bar z_2)=\tau(z_1,z_2)$ and the reverse triangle inequality gives
$\tau(x,z_2) \geq \tau(x,z_1) + \tau(z_1,z_2) = \tau(z_1,z_2)$.
Thus $\tau(x,z) = \tau(z_1,z)$ for all $z\in I^+(z_1) \cap U$.

More generally, if $z \in I^+(x) \cap U$, the argument of Theorem \ref{thm-tlfpcbb-reg} applies. 
\end{pr}

To illustrate curvature bounds in non-regular spaces, consider the following example:
\begin{ex}\label{ex:non-regular}
Consider the following subset of $\R^3_1$: $X=\{(t,v,0):t\leq 0\}\linebreak\cup\{(t,v,t):t\in[0,1]\}\cup\{(t,v,1):t\geq1\}$, with the induced intrinsic $\tau$ (see Figure \ref{fig:non-regular-example}). Then we have a map $f:X\to\R^2_1$ defined by $f(t,v,w)=(t-w,v)$ which collapses the null strip into a spacelike line. This is $\tau$-preserving and monotonic in terms of the causal relation (but not strictly monotonic, so not $\leq$-preserving). $X$ is not regular: the preimage under $f$ of any timelike maximizer $\gamma$ in $\R^2_1$ crossing $\{(0,v)\}$ is the image of a maximizer and will contain a null piece.

Any curvature condition that does not imply any causal relations in $X$ (and that does not assume regularity) will be implied by the curvature bounds on the target of $f$. On the other hand, for any condition implying causal relations one can easily construct counterexamples. Hence
$X$ has timelike curvature bounded above and below by $0$ in the sense of timelike and causal triangle comparison and four-point comparison, but it does not satisfy causal curvature bounded below by $0$ in the sense of strict causal four-point condition and causal curvature bounded above by $0$ in the sense of strict causal triangle comparison. Note that it is not distinguishing, is locally isolating, and has branching of geodesics between timelike related points.

The proof of both Theorem \ref{thm-tlfpcbb-reg} and Theorem \ref{tlcbb-reg} still imply that for the null part $[x,z_1]$ of a non-regular maximizer, for any $z_2\in I^+(x)$ a maximizer between $x$ and $z_2$ goes through $z_1$, and similarly for any $y_2\in I^-(z_2)$ a maximizer between $y_2$ and $z_2$ goes through $x$. But even though these maximizers have a common (null) section, they do not necessarily join to a single maximizer: the preimages under $f$ of any two maximizers in $\R^2_1$ meeting at some $(0,v)$ give rise to this situation.
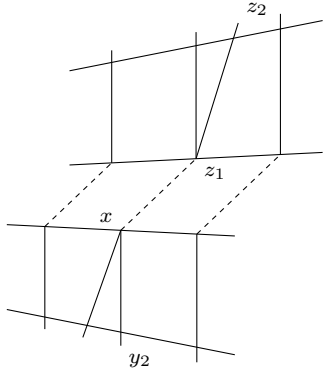
\begin{figure}
    \centering\scriptsize
    \begin{tikzpicture}

        \draw ({10*0.75},{-2*0.75}) -- ({10*1.05},{-2*1.05}); \draw ({10*0.75},{-0.5*0.75}) -- ({10*1.05},{-0.5*1.05}); \draw ({11.1*0.75},{0.555*0.75}) -- ({11.1*1.05},{0.555*1.05}); \draw ({11.1*0.75},{2.22*0.75}) -- ({11.1*1.05},{2.22*1.05}); 
        
        \draw (10,-2.2) -- (10,-0.5)  (11.1,+0.555) -- (11.1,2.42);
        \draw ({10*0.9},{-2.2*0.9}) -- ({10*0.9},{-0.5*0.9})  ({11.1*0.9},{0.555*0.9}) -- ({11.1*0.9},{2.42*0.9});
        \draw ({10*0.8},{-2.2*0.8}) -- ({10*0.8},{-0.5*0.8})  ({11.1*0.8},{0.555*0.8}) -- ({11.1*0.8},{2.42*0.8});

        \draw[dash pattern=on 2pt off 2pt] (10,-0.5) -- (11.1,+0.555) ({10*0.9},{-0.5*0.9}) -- ({11.1*0.9},{0.555*0.9}) ({10*0.8},{-0.5*0.8}) -- ({11.1*0.8},{0.555*0.8});
        \draw ({1*(10*0.85)+0*(10*0.9)},{1*(-2.2*0.85)+0*(-0.5*0.9)}) -- ({10*0.9},{-0.5*0.9}) ({11.1*0.9},{0.555*0.9}) -- ({1*(11.1*0.95)+0*(11.1*0.9)},{1*(2.42*0.95)+0*(0.555*0.9)});
        \draw ({10*0.9},{-2.2*0.9}) node[anchor=north west]{$y_2$} ({10*0.9},{-0.5*0.9}) node[anchor=south east]{$x$} ({11.1*0.9},{0.555*0.9}) node[anchor=north west]{$z_1$} ({1*(11.1*0.95)+0*(11.1*0.9)},{1*(2.42*0.95)+0*(0.555*0.9)}) node[anchor=south west]{$z_2$};
    \end{tikzpicture}
    \caption{The space $X$ from Example \ref{ex:non-regular}. Two domains isometric to a Minkowski half-space are joined by a null strip. The maximizers $[y_2, z_1]$ and $[x,z_2]$ overlap in their null sections but do not combine to give a maximizer.}
    \label{fig:non-regular-example}
\end{figure}
\end{ex}

\section{Inextendibility versus unboundedness of curvature}
We now apply the results of Section \ref{sec-reg} to the question of inextendibility of \LpLSn s or spacetimes. First, we recall and slightly generalize the notion of an extension of a \LpLS as used in \cite{GKS:19}.

\begin{defi}[Extension]
An \LpLS $(\tilde{X}, \tilde{\ell})$ is an extension of \Xll if:
\begin{itemize}
    \item $\tilde{X}$ is connected.
    \item There is a homeomorphism $\iota\colon X \rightarrow \tilde{X}$ onto a proper open subset $\iota(X)$.
    \item The map $\iota$ preserves the time separation function, i.e., for all $x,y\in X$ we have $\tilde\ell(\iota(x),\iota(y)) = \ell(x,y)$.
\end{itemize}
\end{defi}

One could always extend a \LpLS by adding points on the \emph{spacelike boundary}, i.e., all points $\tilde x$ such that $\tilde\ell(\tilde x, \iota(x)) = \tilde\ell(\iota(x),\tilde x) = -\infty$, as introduced in \cite{MS:24}. If the space is future or past distinguishing, then there can be only one point in the spacelike boundary as then trivially $I^\pm(x)=\emptyset = I^\pm(y)$. Moreover, being not locally timelike isolating implies that there is no spacelike boundary.
\medskip

Next, we recall the \emph{TC}-condition from \cite{GKS:19} that works as an analogue of timelike geodesic completeness: A \LpLS \Xll satisfies the \emph{TC}-condition if every inextendible timelike maximizer has infinite length. By \cite[Lem.\ 5.2]{GKS:19} a smooth and strongly causal spacetime is timelike geodesically complete if and only if it satisfies the \emph{TC}-condition.

\begin{thm}[Inextendibility versus regularity]
Let \Xll be a Lorentz\-ian pre-length space satisfying the \emph{TC}-condition. Then $X$ is inextendible as a regular and weakly normal  \LpLS without spacelike boundary.
\end{thm}
\begin{pr}
Assume a regular and weakly normal extension $\iota\colon (X,\ell)\rightarrow(\tilde X,\tilde\ell)$ without spacelike boundary exists. By assumption $\partial\iota(X)\neq \emptyset$, let $\tilde p\in\partial\iota(X)$ and let $\tilde U$ be a weakly normal neighbourhood of $\tilde p$. As $\tilde U$ is weakly normal, let $\tilde r\in I^+(\tilde p)\cap \tilde U$. Consequently, $\tilde V:=I^-(\tilde r) \cap \tilde U$ is an open neighbourhood of $\tilde p$. First, assume that $\tilde r\not\in \iota(X)$ and let $\tilde q = \iota(q)\in \tilde V \cap \iota(X)$. By assumption there is a causal maximizer $\tilde{\gamma}\colon[0,1]\rightarrow \tilde X$ from $\tilde q$ to $\tilde r$, that leaves $\iota(X)$ at a boundary point $\tilde{w}=\tilde\gamma(t)$ for the minimal $t\in(0,1]$. Moreover, by regularity and since $\tilde q\, \tilde \ll\, \tilde r$ we know that $\tilde \gamma$ is timelike. The curve $\gamma = i^{-1} \circ \tilde{\gamma}\rvert_{[0,t)}$ is timelike and inextendible in $X$. However, its length is:
\begin{align*}
 L_{\tau}(\gamma) &=  \lim_{s \nearrow t} L_{\tau}(\gamma_{\rvert_{[0,s]}}) =  \lim_{s \nearrow t} \tau(q, \gamma(s)) = \lim_{s \nearrow t} \tilde{\tau}(\tilde q, \tilde{\gamma}(s)) \leq \tilde{\tau}(\tilde q, \tilde{w})\\
 &\leq \tilde\tau(\tilde q,\tilde r) < \infty\,,
 \end{align*}
where  we used the reverse triangle inequality to see that $s\mapsto \tilde\tau(\tilde q,\tilde\gamma(s))$ is monotonically increasing and bounded by $\tilde\tau(\tilde q,\tilde w)$. This now contradicts the \emph{TC}-condition. Second, if $\tilde r = \iota(r)\in \iota(X)$ we can find $\tilde q \in \tilde V \cap \tilde X\backslash \iota(X)$. Now a time reversed version of the argument above yields the desired contradiction.
\end{pr}

In particular we have that
\begin{cor}
 A strongly causal smooth spacetime (or \LLSn) that is timelike geodesically complete (i.e., satisfies the \emph{TC}-condition) is inextendible as a regular \LLSn.
\end{cor}

Finally, this means that there cannot be an extension which has causal curvature bounded below by Theorem \ref{thm-ccbb-reg} (or Corollary \ref{cor-tlfpcbb-reg}). Note, that weakly normal implies that there is no spacelike boundary.
\begin{thm}
 A locally distinguishing \LpLS satisfying the \emph{TC}-condition cannot be extended as a locally distinguishing weakly normal \LpLS having causal curvature bounded below  (and $\tau$ locally continuous in the case of 4-point TLCBB).
\end{thm}

In particular, this applies to all strongly causal timelike geodesically complete spacetimes. 
\bigskip

{\footnotesize
{\bf\noindent Acknowledgments}\\
\noindent We are grateful to Michael Kunzinger, Karim Mosani, Lewis Napper, Raquel Perales, Jona R\"ohrig, Felix Rott, Kharanshu Solanki, Roland Steinbauer and Inés Vega González for stimulating and helpful discussions. The anonymous reviewers provided important and useful feedback which strengthened the work, for which we thank them. This research was funded in whole or in part by the Austrian Science Fund (FWF) [Grant DOIs \href{https://doi.org/10.55776/STA32}{10.55776/STA32}, \href{https://doi.org/10.55776/J5001}{10.55776/J5001}].
JH was supported by a UKRI Future Leaders Fellowship [grant number MR/W01176X/1].
Part of this work was carried out while JH was a guest of the University of Vienna supported by the Austrian Science Fund (FWF) [Grant DOI \href{https://doi.org/10.55776/EFP6}{10.55776/EFP6}].
For open access purposes, the authors have applied a CC BY public copyright license to any author accepted manuscript version arising from this submission.}

{\footnotesize 
 \phantomsection
 \addcontentsline{toc}{section}{References}

\bibliographystyle{halpha-abbrv}
\bibliography{Master}
}

\end{document}